# A Novel Labeled Human Voice Signal Dataset for Misbehavior Detection

**Ali Raza**
Department of Software Engineering
The University Of Lahore
Lahore 54000, Pakistan
ali.raza.scholarly@gmail.com

**Faizan Younas**
Department of Computer Science &
Information Technology
The University Of Lahore
Lahore 54000, Pakistan
younasfaizan97@gmail.com

## ABSTRACT

Voice signal classification based on human behaviours involves analyzing various aspects of speech patterns and delivery styles. In this study, a real-time dataset collection is performed where participants are instructed to speak twelve psychology questions in two distinct manners: first, in a harsh voice, which is categorized as "misbehaved"; and second, in a polite manner, categorized as "normal". These classifications are crucial in understanding how different vocal behaviours affect the interpretation and classification of voice signals. This research highlights the significance of voice tone and delivery in automated machine-learning systems for voice analysis and recognition. This research contributes to the broader field of voice signal analysis by elucidating the impact of human behaviour on the perception and categorization of voice signals, thereby enhancing the development of more accurate and context-aware voice recognition technologies.

## 1. Introduction

### 1.2 Voice linguistics

Voice linguistics is a subfield of linguistics that examines the role of voice in language [1], encompassing both the physiological and acoustic properties of vocal sound production and their linguistic implications. Voice linguistics involves analyzing the mechanisms of phonation, the variations in vocal fold vibration, and the acoustic features of the resultant sound waves. Voice linguistics also investigates sociolinguistic aspects, such as how voice characteristics can signal identity, emotion, and social status, and how they vary across different languages, cultures, and contexts. Through interdisciplinary approaches, voice linguistics aims to deepen our understanding of the intricate relationship between voice and language.

### 1.2 Voice emotion signal analysis

Voice emotion signal analysis is a field of study that involves examining the acoustic properties of a person's voice to identify and interpret their emotional state. This analysis typically utilizes machine learning algorithms and signal processing techniques [2, 3, 4] to detect subtle variations in pitch, tone, intensity, and speech rate that correlate with different emotions such as happiness, sadness, anger, or fear. The need for voice emotion signal analysis arises from its wide range of applications in various domains.

In healthcare, it can aid in diagnosing and monitoring mental health conditions by providing objective measures of emotional well-being. In customer service, it can enhance user experience by enabling more empathetic and responsive interactions. Additionally, human-computer interaction allows for the development of more intuitive and adaptive systems that can respond appropriately to the emotional states of users. Overall, voice emotion signal analysis holds significant potential for improving communication, understanding human emotions, and creating more personalized and effective technological solutions.

### 1.3 Applications of voice misbehaviour

Voice signal analysis has a wide range of applications, particularly in identifying and addressing voice misbehaviour, which refers to abnormal or improper use of the voice. These applications are critical in various fields [5, 6, 7], including healthcare, security, and human-computer interaction. In healthcare, voice signal analysis can be employed to diagnose and monitor conditions such as vocal fold nodules, polyps, or other disorders that affect vocal quality. It can also be used to detect early signs of neurological diseases like Parkinson's, where voice changes can be an early indicator.

In the realm of security, voice signal analysis is utilized for speaker verification and identification, helping to prevent fraud and enhance security systems. This technology can detect stress or deception in a person's voice, which is useful in lie detection and security screenings. Furthermore, in human-computer interaction, voice signal analysis improves speech recognition systems [8, 9, 10], enabling more accurate and natural interactions with virtual assistants and automated systems. The development of sophisticated algorithms and machine learning techniques has significantly enhanced the ability to analyze and interpret voice signals, making it a powerful tool for detecting and managing voice misbehaviour across various applications.

### 1.4 Research Data Objectives

- This area of study explores how different voice qualities, such as pitch, loudness, and timbre, are utilized in communication and how they contribute to meaning beyond the lexical and grammatical components of language.

- For understanding human behaviour using real-time voice signal analysis, this study focuses on the detection of emotions embedded within vocal expressions. By collecting voice signals in real time, we aim to capture the nuances and subtleties that characterize emotional states such as happiness, sadness, anger, and fear.

## 2 Experimental Design, Materials and Methods

Our research proposed experimental design, materials and methods are examined in this section. A systematic approach is employed to classify human speech into two distinct categories: normal and misbehaving signals.

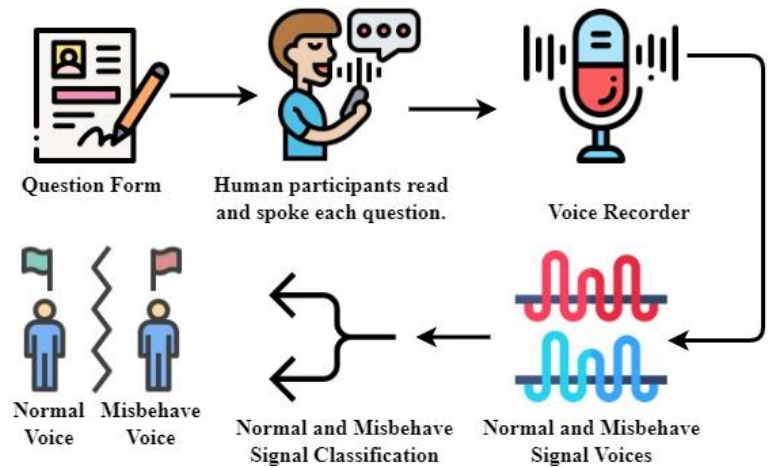

Figure 1: The proposed real-time data collection methodology workflow.

### 2.1 Data Collection Design

In this study, a systematic approach is employed to classify human speech into two distinct categories: normal and misbehaving signals as shown in Figure 1.

- Initially, each participant was given a questionnaire consisting of 12 questions.

- Each participant read and spoke each question aloud, and their responses were recorded using a high-fidelity voice recorder.

- The recorded audio signals were then analyzed and classified into two categories: normal and misbehaving signals.

This classification aimed to differentiate between standard, expected speech patterns and those that exhibited unusual or disruptive characteristics. The process involved both qualitative and quantitative analysis, utilizing advanced signal processing techniques to accurately identify and categorize the recorded speech patterns. This research provides insights into the potential for automated systems to detect and classify human speech based on behavioural cues.

The following twelve questions were spoken by each participant during the data collection:

- Could you please lower your voice?

- I can't believe you just said that.

- Excuse me, I need your attention.

- That's not appropriate behaviour.
- Could you repeat that, please?
- I appreciate your cooperation.
- This is unacceptable behaviour.
- You need to calm down right now!
- Stop being so rude!
- I don't appreciate your tone.
- Could you explain your point of view?
- Thank you for your cooperation.

## 2.2 Voice Signal Classification

Voice signal classification based on human behaviours involves analyzing various aspects of speech patterns and delivery styles. In a study focusing on this, participants are instructed to speak a question in two distinct manners: first, in a harsh voice, which is categorized as "misbehaved"; and second, in a polite manner, categorized as "normal". These classifications were crucial in understanding how different vocal behaviours affect the interpretation and classification of voice signals [11, 12], highlighting the significance of tone and delivery in automated systems for voice analysis and recognition.

## 2.3 Data Privacy

Data privacy and the protection of participants' personal information are of utmost importance in this study. All real-time voice signals collected to detect behaviours based on emotions within the voice are handled with strict confidentiality. Participants are informed about the nature and scope of the data collection process, and their explicit consent is obtained before any data acquisition. Voice recordings are anonymized to ensure that individuals cannot be identified from the data.

## 3 Conclusion and future work

This study performed a real-time dataset collection where participants were instructed to speak twelve psychology questions in two distinct manners: first, in a harsh voice, which is categorized as "misbehaved"; and second, in a polite manner, categorized as "normal". These classifications are crucial in understanding how different vocal behaviours affect the interpretation and classification of voice signals. This research highlights the significance of voice tone and delivery in automated machine-learning systems for voice analysis and recognition.

## 3.1 Future Work

In the future, we plan to develop advanced machine learning and deep learning techniques utilizing this newly collected dataset for human misbehave detection through voice signals. Additionally, we will augment the dataset through further collection efforts, enhancing its richness and utility for innovative analytical applications.

## Ethics statements

The study was conducted in accordance with the Fareed Computing and Research Center, KFUEIT. The data collection adhered to relevant regulations and guidelines. The authors are authorized to publicly share the data.

## CRediT Author Statement

Ali Raza: Conceptualization, Software, Validation, Data curation, Writing – original draft, Supervision, Project administration; Faizan Younas: Software, Validation, Data curation, Supervision, Project administration; Ali Raza: Methodology, Investigation, Resources, Supervision; Writing – original draft, Validation, Data curation.

## Declaration of interests

The authors declare that they have no known competing financial interests or personal relationships that could have appeared to influence the work reported in this paper.

## Data Availability

The data used in this study is available upon request. Interested researchers can contact Ali Raza to request access to the data. Requests will be reviewed and evaluated in accordance with the ethical and legal guidelines governing data sharing.

## Acknowledgments

We are deeply grateful to all those who contributed to this article. Co-authors would like to thank Ali Raza for his insights and expertise, which are instrumental in shaping the direction of this research process.